# Measurement of Investment activity in China based on Natural language processing technology[1]

Xiaobin Tang[a], Tong Shen[a*] and Manru Dong[b]

[a]*School of Statistics, University of International Business and Economics, Beijing, China*

[b]*School of Information, Beijing Wuzi University, Beijing, China*

[*]*Corresponding author at: School of Statistics, University of International Business and Economics, Beijing, 100029, China.*

*E-mail: shen_tong01@163.com*

**ABSTRACT:** The purpose of this study is to propose a new index to measure and reflect China's investment activity in time, and to analyze the changes of China's investment activity in the past five years. This study first uses the NEZHA model for semantic representation, and expand the indicator system based on semantic similarity. Then we calculate China's investment activity index by using the network search data. This study shows that China's investment activity began to decline in 2019, rebounded for a period of time after the outbreak of COVID-19 in 2020, and then continued to maintain a downward trend. Private investment activity has declined significantly, while government investment activity has increased. Among the provinces in Chinese Mainland, the investment activity of economically developed provinces has decreased significantly, while the investment activity of some economically less developed provinces in the north and south is higher. Our research will provide timely investment information for the government, decision makers and managers, as well as provide other researchers who also pay attention to investment with a perspective other than investment in fixed asset.

**KEYWORDS:** Investment Activity Assessment; Keywords Expansion; NEZHA

---

[1] We thank MindSpore for the partial support of this work, which is a new deep learning computing framwork\footnote{https://www.mindspore.cn/}.

Model; Entropy Weight

# 1. Introduction

Investment, consumption and export are known as the "three carriages" driving China's economic growth. Consumption is greatly affected by domestic income, debt and domestic economic environment, while export is not only affected by the domestic economy, but also by the economic development of other countries, cross-border logistics and international relations. Investment, by contrast, is more regulated by the government and more useful in times of economic distress. While private and foreign investment are highly affected by the economic environment, the government is often able to defuse the crisis and boost the economy through large infrastructure investments. At present, the economic environment caused by the global epidemic of COVID-19 and the impact of international relations is not optimistic. The decline of consumption and the instability of import and export make China's economy face great downward pressure. Therefore, it is of great significance to timely study the investment measurement and analyze the investment activity of various provinces for China's economic development.

Traditional investment-related indicators are delayed and low-frequency due to statistical methods, which reduces the timeliness of investment activity measurement. Considering that people often conduct Internet search to obtain information when they have investment ideas and then make investment decisions, it shows that Internet search data reflects people's active degree in investment ideas. Natural language processing technology can obtain search information, and network search data can better overcome the shortcomings of traditional data. Therefore, this paper considers the application of natural language processing technology and network search data in the study of investment indicators, and timely measure the investment situation.

The structure of this paper is as follows: Chapter two reviews the previous studies; The third chapter introduces the model used in this paper. The fourth chapter introduces the construction process of China's investment activity index; Chapter five calculates the active degree of Chinese investment through empirical research. Chapter six is the conclusion.

**2. Literature review**

With the development of internet big data, the internet search index has received lots of attention of researchers. There is a large volume of published studies using web search data. Google trend is the representative of such data, and the research on the use of Google trend first appeared in 2009 (Jun et al., 2018). The research published in Nature by Google researchers used google trends to track influenza diseases in the population (Ginsberg et al., 2009), and Google researchers used google trends to predict economic indicators such as unemployment rate and consumer confidence (Choi and Varian, 2009), and found that google trends were related to house and car sales in subsequent studies (Choi and Varian, 2012). In the economic and financial fields, search engine data also has many applications. Researchers compared the search index with the consumer confidence index and the University of Michigan consumer sentiment index (MCSI), and found that the prediction ability of search engine data may be higher than the survey index (Vosen and Schmidt, 2011). Da et al. Proposed Google search volume index (SVI) to measure investor interest, and used SVI to predict stock prices and initial IPO profits (DA et al., 2011). Preis et al. Proposed the stock market early warning method using the search volume of financial words (Preis et al., 2013), Silva et al. used Google trend to analyze the behavior of fashion consumers and predict the trend of fashion consumers (Silva et al., 2019). Previous studies have shown that web search data can play an important role in the economic and financial fields, and also provide theoretical and practical support for this paper to use web search data to carry out research.

Natural language processing(NLP) technology is widely used in machine translation, viewpoint extraction, text classification, answering questions and so on. It integrates linguistics, computer science and mathematics. The development of NLP methods has gone through the process from statistical model to neural network model. NLP model based on statistical methods usually calculates the probability of word occurrence by word frequency. The representative method is TF-IDF algorithm (Salton and Yu, 1974) which uses word frequency to replace probability to extract keywords. PageRank algorithm (page et al., 1999) is an algorithm used in Google search. It judges the importance through links, so as to sort the search results according to the importance.Textrank algorithm (mihalcea and tarau, 2004) is improved from PageRank algorithm. It uses the co-occurrence semantics of words in the document to extract keywords, keyword groups and key sentences from a given document. LDA model (BLEI et al., 2003) is used to generate the subject of an article and can give the subject of an article in the form of probability distribution. The natural language processing model based on neural network uses neural network modeling to determine the parameters in the network through the training of the model, which can learn and mine the deeper semantic information in the text. Word2vec algorithm (mikolov et al., 2013) is based on shallow neural network and can express words as word vectors, which is widely used. The recurrent neural network language model can obtain long context semantic information and has strong semantic representation ability; After 2018, the pre-training language model based on transformer and Bert model (Vaswani et al., 2017; Devlin et al., 2018) has been greatly welcomed. This kind of model adopts self-monitoring training method and uses a large number of corpus for training, and has achieved the best results in a series of downstream tasks; In terms of the Chinese model, NEZHA model (Wei et al., 2019) which is improved from Bert model, it considered the characteristics of the Chinese language, and has made achievements beyond the Bert-Chinese model in the Chinese downstream tasks.

In terms of the application of NLP methods, many studies have used NLP technology to conduct stock forecasting research. Due to the short text length and rich information, twitter text has received the attention of researchers (Bollen et al. 2011; Si et al. 2013; Wei et al. 2016). Xiong Kai and others used words, triples composed of subjects, verbs and objects, and sentence extraction information to predict the stock market (Xiong et al., 2021).

Through reviewing previous studies, it can be found that although the research on NLP and web search data has developed rapidly, there has been little discussion about following points: first, the researchs using text data are mainly focused on the prediction of stock market and foreign exchange rate (Xing et al., 2018), and very little attention has been paid to economic indicators and economic measurement. Second, previous studies often only used the method of extracting specific keywords, key phrases or emotional analysis to extract semantic information, without learning the deep semantics of the text. With big data from internet, still using simple text mining methods will lead to a waste of information and reduce the effect of information mining and acquisition. Third, in the past, when using search data, researchers often directly selected the corresponding keywords of the research object, or only selected keywords from several aspects. There was no system to select words from more dimensions, and the factors affecting the research object at all levels could not be included in the research scope, resulting in the reduction of effective information acquisition.

Therefore, based on the above shortcomings, this paper proposes the following research methods: firstly, in view of the analysis of Chinese investors' Internet search behavior, this paper constructs the indicators system of Chinese investment activity. Secondly, we obtain the relevant text of the corresponding indicators from internet as corpus to train the Nezha model; Thirdly, we segment these text to obtain extended candidate words, and then by using Nezha model these words are represented as vectors. The extended words are selected through the correlation of word vectors to achieve the improvement of the indicator system. Finally, we use Baidu search index of expanded indicator system to calculate China's investment activity.

## 3. Methodology

In this paper we use web search data, namely Baidu Index data as the data of our indicators. We chose Baidu Index because in China Baidu is the most commonly used search engine and Baidu Index is similar to Google trend. Unlike traditional economic index, one web search word may have many other similar words, and the meaning is not clear enough. In order to prevent the influence of subjective factors that may be caused by the artificial selection of indicators or the influence of missing information caused by the limited understanding of researchers, this paper uses the NEZHA model to expand the artificially selected indicators. NEZHA model is a Chinese language pre- training model developed by Huawei Noah laboratories based on the Bert model and improved according to Chinese characteristics. Next, we introduces NEZHA model.

NEZHA model also have two parts. The first part is the input representation, that is, the embedding process, and the second part is the stacking of multiple encoders. The purpose of embedding process is to convert language into vectors, and use different vectors to represent the information contained in different words. In NEZHA model, embedding consists of three parts: token embedding, segment embedding and position embedding, as shown in Figure 1. Among them, token embedding represents words as vectors; segment embedding represents the relationship between sentences. Words in the same sentence have the same segment embedding; The position embedding is used to express the order between words. Because NEZHA model inputs multiple sentences at the same time instead of words in order,unlike other NLP models the location information needs to be added to represent the order information. In the part of token embedding, NEZHA model also specially add two symbols [cls] and [sep], which are placed at the beginning of the text and at the place where the text breaks, so as to use the output of the two symbols to judge whether there is a context relationship between sentences in subsequent training. This setting also makes the output of [cls] symbols after training have the ability to represent sentence semantics. After three kinds of embedding, NEZHA model finally uses the sum of three embedding vectors as input.

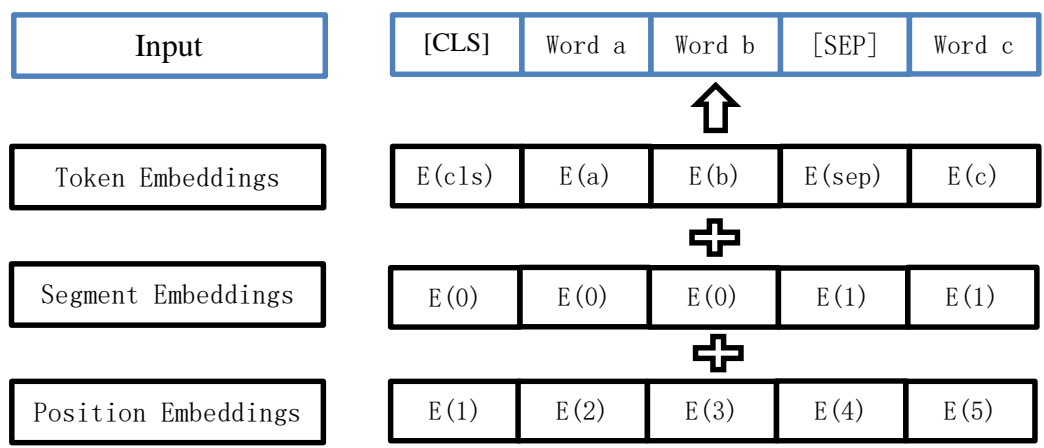

Figure 1. Three kind of Embedding in NEZHA

The second part of NEZHA is composed of multiple encoders. Different versions of NEZHA use different numbers of encoders. Each encoder is mainly composed of multi-head attention and feed forward networks. The output results of each part need to be added and normalised before entering the next step, as shown in Figure 2.

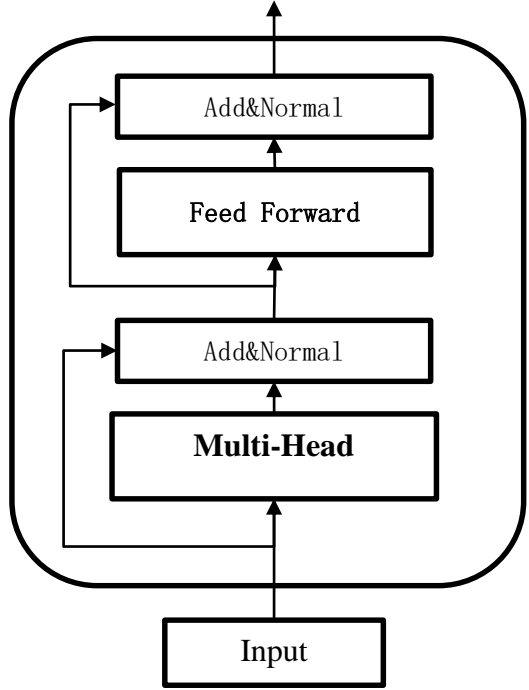

Figure 2. Encoder architecture of NEZHA

Multi-head attention is the most important part of the encoder, and it is also the most critical part of semantic extraction. Multi-head attention is composed of multiple self-attention. Each self-attention can obtain semantic information from different dimensions, and the multi-head attention concentrates the information of each head and outputs it. After that, the data needs to pass through a layer of residual sum and normalization process (Add&Norm) ,than a layer of feed-forward neural network (FFN), which contains two linear transformation and one relu activation function:

$$FFN(x)=\max(0, xW_1+b_1)W_2 + b_2 \qquad (1)$$

Finally, after a layer of residual sum and normalization process (Add&Norm), the output of an encoder is completed.

Compared with Bert model, the Nezha model has been improved in four parts: first, in position embedding, Transformer model and Bert model both use absolute position embedding (APE), while NEZHA model takes into account the relative position relationship of the text. The addition of fully functional relative position embedding (RPE) to multi-head attention improved the performance of NEZHA model in downstream tasks. Specifically, if the sentence is input as $x = (x_1, x_2, \cdots, x_n)$, The single self attention value of word $x_i$ is:

$$Attention(q_i, k_i, v_i) = soft\max(\frac{q_i(k_i + \alpha_{ij}^k)^T}{\sqrt{d_k}})(v_i + \alpha_{ij}^v) \quad (2)$$

$\boldsymbol{\alpha}_{ij}^v$ and $\boldsymbol{\alpha}_{ij}^k$ are the vectors representing the relative position embedding between position I and position J. In previous studies (Shaw et al., 2018), the relative position embedding containing parameters is often used, and the parameters need to be learned in the pre-training process, while the NEZHA model uses functional embedding, so there is no parameter to learn. The embedding method is as follows:

$$\alpha_{ij}[2k] = \sin((j-i)/10000^{\frac{2 \cdot k}{d_z}}) \quad (3)$$

$$\alpha_{ij}[2k+1] = \cos((j-i)/10000^{\frac{2 \cdot k}{d_z}}) \quad (4)$$

Second, in pre-training, NEZHA model considers the existence of Chinese words and uses the whole word masking (WWM) technology. A word in Chinese is often composed of multiple tokens, while Bert-Chinese model only covers a single token. In NEZHA model, once a token is covered, the whole word will also be covered. Such improvement is conducive to model learning and understanding of the meaning of words. Third, the NEZHA model adds a span prediction task in pre-training, which will cover a continuous span, this task can significantly improve the effect of the downstream span extraction task. Fourth, in training NEZHA model adopts mixed precision training. In the traditional deep learning training, all variables are in fp32 (single precision floating point number) format, while the mixed precision training converts parameters to fp16 (half precision floating point number) format in forward and backward propagation, which reduces the space occupied by parameters and improves the calculation speed. Fifth, the layer-wise adaptive moments optimizer (LAMB) for batching training is adopted. In deep learning training, when the batch size is large, the learning rate needs to be adjusted, otherwise the model performance may be greatly affected. The LAMB optimizer can adaptively adjust the learning rate so that the model does not lose performance in mass training, and greatly shortens the model training time.

## 4. An index for evaluation of investment activity in China

### *4.1. Indicator system for evaluation of investment activity in China*

A reasonable indicator system is the fundation of our measurment. In order to reflect the search behavior of Chinese investors after they generating investment ideas, it is necessary to select key words related to investment behavior to build an indicator system. This paper divides words based on investment subject and investment environment, and analyzes the main factors related to investment.

(1) Government investment

Government investment is an important part that distinguishes China from other countries. Especially when the economy is under downward pressure, the government often achieves the purpose of stimulating economic growth and maintaining social stability through huge amount of investment. In the government investment, this paper selects the following three dimension as representatives: (a) government economic attitude. The changes in the range of government investment will be reflected through relevant meetings and policies, so this paper selects the terms "central economic work conference" and "local debt" as the indicator words at this dimension. (b) Infrastructure investment. Infrastructure investment is the main way for the Chinese government to invest. In 2008, in response to the impact of the international financial crisis, the Chinese government launched ten measures with a total investment of about 4trillion yuan to further expand domestic demand and promote steady and rapid economic growth. Of the 4trillion yuan, more than 1.5 trillion yuan is used for infrastructure construction, excluding the 1trillion yuan invested in earthquake reconstruction. It can be seen that infrastructure investment is the key element of government investment. This paper selects the main construction objectives included in infrastructure construction as the indicator words of infrastructure investment. (c) New infrastructure investment. Traditional infrastructure investment is still important, but it can not meet the needs of China's economic development. Therefore, new infrastructure construction has become a hot spot of current government investment. The new infrastructure mainly includes, for example, 5g base stations, UHV, charging piles for new energy vehicles, big data centers, artificial intelligence and industrial Internet. Facing high-quality economic development, it provides services such as transformation and upgrading, integration and innovation. This paper refers to the business field of Shenzhen new infrastructure 50 index sample stock company launched by Shenzhen Stock Exchange, and selects the indicator words.

(2) Investment influenced by government

The investment behavior of general enterprises is for the development of the company and making profits. However, the investment behavior of many Chinese companies, such as state-owned enterprises, central enterprises and public institutions, is to comply with the policy requirements or the development of industries supported by the state, not simply for short-term economic interests. Since state-owned enterprises and central enterprises are generally large enterprises, this part of investment also needs to be considered. In this part, this paper selects the corresponding enterprise words as indicators.

(3) Private investment

Private investment is the main part of China's investment. In China's fixed asset investment, private investment generally accounts for about 60%. Considering the special status of real estate investment in China's fixed asset investment, this paper divides private investment into real estate investment and non real estate investment. (a) Real estate investment. Real estate investment accounts for nearly 50% of China's fixed asset investment, which is of great significance to China's economy. Rising house prices will lead residents to speculate in real estate and damage the real economy, while falling house prices will trigger a chain reaction and have a negative impact on the economy. In recent years, the Chinese government has issued a series of policies to regulate the real estate market and maintain the stability of house prices. In the field of real estate investment, this paper selects real estate and house price related words as indicators. (b) Non real estate investment. In this paper, private investment except real estate investment is divided into this kind of investment. In terms of indicators, we take into account the sites and processes required by private entities when making investment, and selects the corresponding words.

(4) Foreign investment

Foreign investment is an important part of China's investment. Although there is a large difference between foreign investment and domestic investment in terms of investment amount, in the case of global epidemic and inflation, China's excellent epidemic control, complete supply chain and low inflation rate may become important factors to attract world investment, especially long-term investment. Therefore, foreign investment is also the research object of this paper. In terms of foreign investment, we select words relate to foreign direct investment as indicators.

(5) Investment environment

Investment behavior is closely related to the economic environment. Liquidity, interest rates and industrial policies will have a great impact on investment behavior. This paper divides the investment environment into two levels: (1) Industry aspect. At this aspect, this paper mainly selects the words related to the real economy to represent the environment of the real economy industry. (2) Economic aspect. In terms of economic environment, this paper mainly selects words related to interest and tax to represent the environment of investment funds.

Based on the analysis of China's investment activity, this paper sets out from the five dimensions of government investment, Investment influenced by government, private investment, foreign investment and investment environment, and on the basis of giving consideration to comprehensiveness, representativeness and availability, constructs an indicator system of five primary indicators, nine secondary indicators and 47 indicator entries, as shown in Table 1.

Table 1. Indicator system for measure investment activity in China

| Primary indicators | Secondary indicators | Indicator entries |
|---|---|---|
| Government investment | Government economic attitude | Central economic work conference |
| | | Local debt |
| | Infrastructure investment | Infrastructure |
| | | Affordable housing |
| | | Highway |
| | | Railway |
| | | Airport |
| | | Electrified wire netting |
| | | Freeway |
| | | Energy saving and emission reducti |
| | | Livelihood project |
| | | Water conservancy |
| | New infrastructure investment | New infrastructure |
| | | 5G |
| | | UHV |
| | | New energy vehicles |
| | | Industrial Internet |

| | | Artificial intelligence |
| --- | --- | --- |
| | | Data center |
| Investment influenced by government | State-owned enterprise investment | State-owned enterprise |
| | | Central enterprises |
| | | Public institutions |
| Private investment | Real estate investment | Real estate |
| | | Housing price |
| | | Real estate policy |
| | Non real estate investment | Limited company |
| | | Private enterprise |
| | | Self-employed laborer |
| | | Enterprise Registration |
| | | Office Building |
| | | Shops |
| | | Factory |
| | | Workshop |
| Foreign investment | Foreign investment | Fdi |
| | | Foreign enterprise |
| | | Joint venture |
| | | The open policy |
| Investment environment | Industry aspect | Real economy |
| | | Manufacturing |
| | | Industry |
| | | Small and micro businesses |
| | | Industrial upgrading |
| | Economic aspect | Tax reduction |
| | | Interest rate cut |
| | | Reserve requirement ratio cut |
| | | Loan |
| | | Interest rate |

## 4.2. Expand indicators based on NEZHA model

After selecting the indicators manually, we expand the indicators according to the semantic similarity. First, this paper collects the corpus related to each indicator, and then uses Jieba word segmentation to divide the words in the corpus and remove the stop words. The original indicator words are used as seed words and the remaining words become expansion candidates. Then we use collected corpus to train the NEZHA model, so we can improve the performance of NEZHA model in investment related corpus. Finally,

NEZHA model is used to vectorise the candidate words and the seed words, and the cosine values between the vectors are calculated as similarity to select the candidate words that are most similar to the seed words, so as to expand the indicators. In terms of corpus acquisition, considering the quality of the corpus and the degree of correlation with the indicators, this paper selects the Baidu Encyclopedia entries related to the indicators as the corpus, which is also the same as the corpus used in the pre-training of NEZHA model. In details, we set the expansion candidate words of seed words contained in each secondary indicators, and selects the words with the highest similarity ranking of 50 as the expansion words. Further, through manual discrimination, adjectives, adverbs and words with unclear meaning are removed. Since not all words have Baidu search index, this paper only retains the words with Baidu Index as the final expansion words. The specific process is shown in Figure 3.

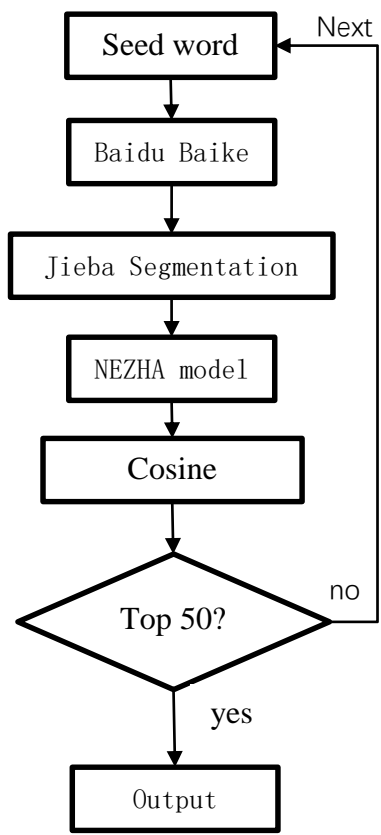

Figure 3. Words similarity selection process

*4.2.1. Pre-training of NEZHA model*

Like other large-scale deep learning models, NEZHA model needs pre-training with massive data before use, so that the model can have the ability to express rich semantic information. NEZHA model uses Chinese Wikipedia, Baidu Encyclopedia and Chinese News Texts in the pre-training. In terms of training tasks, the masked language model (MLM) task and next sentence prediction (NSP) task are mainly executed. Specifically, the MLM task is to randomly cover some words in the sentence, and the model needs to predict the covered part. MLM training can use the information of the uncovered part of the context. Therefore, unlike the previous one-way language model, Bert model has the ability of two-way language representation. The NSP task is to let the model judge whether sentence B is the next sentence of sentence A when sentence A and B come together. NSP task enables the model to understand the relationship between sentences, while MLM task focuses on semantic understanding at the word level. Finally, the model takes the loss sum of each training task as the training goal.

*4.2.2. NEZHA model fine-tuning and text similarity calculation*

Pre-trained Nezha model has a good language representation ability. However, in order to further improve the effect of the model on investment related texts, it is also necessary to use such texts to train the model, that is, fine-tune. The purpose of this paper is to enable the model to learn the semantic similarity between seed words and extended candidate words, so the downstream task of this paper is classification that to determine whether the keywords and sentences are related. In terms of corpus, in addition to the Baidu Encyclopedia text, this paper also uses the tnews data set in the CLUE data set. The CLUE dataset is a benchmark dataset for Chinese language understanding evaluation. For English texts, there are GLUE, supeGLUE and other datasets as benchmarks for the effect evaluation of various pre-training models. For Chinese texts, the CLUE dataset is the first large-scale Chinese language model evaluation benchmark, including 9 datasets for different tasks. The tnews dataset is used for short text classification tasks.

For Baidu Encyclopedia text, we construct the data in [tag, word, text] format, and the tag represents whether the word is the key word of text. If the word and text are related, take 1; otherwise, take 0. For tnews data, this paper selects the part that contains indicator words in the data set, and constructs the data in the same format. Finally, this paper constructs a training set containing 4000 pieces of data and a validation set containing 1000 pieces of data, in which the ratio of label 1 and 0 is 1:1. Use training set and validation set to fine-tune NEZHA model and adjust model parameters. This training actually uses the output vector of the [cls] tag at the beginning of the sentence to calculate the probability of each category through linear transformation and normalization for classification.

Table 2. Hyperparameters for NEZHA fine-tunning

| Parameter | Train | Dev |
| --- | --- | --- |
| Batch Size | 16 | 16 |
| Epoch | 10 | 10 |

After that, the extended seed words and candidate words obtained from Baidu Encyclopedia text segmentation are input into the model again, the [cls] output vector representing the meaning of each word, and we calculate cosine coefficient between vectors of seed words vector and candidate words as word similarity.

### 4.3. Calculate investment activity index in China

After the expansion of the indicator system, we collect Baidu search index of the corresponding indicator entry according to the expanded indictor system, and use the entropy weight method to calculate the indicator weight, so as to calculate the China investment activity index and the activity index in all dimensions.

# 5. Analysis of investment activity mesurement in China

## 5.1. Research dataset

According to the indicator system, this paper selects 47 seed words related to investment activity from each dimension, and collects the daily Baidu search index of indicator words from June 2017 to may 2022. In terms of text, a total of 84214 words and 4545 sentences were collected from Baidu Encyclopedia. After Jieba word segmentation and removal of stop words, 11580 expansion candidates were obtained. In addition, the national fixed asset investment data used in this paper are from the website of the National Bureau of statistics.

## 5.2. Indicator system after expanded

Based on pre-trained NEZHA model, this paper uses corpus that consist of Baidu Encyclopedia and CLUE dataset to fine tune. Because of the difference of version, we train NEZHA model based on pytorch. And we also provide the code of this paper under mindspore. After 10 rounds of fine tuning training, the accuracy of NEZHA model on the validation set increases from 77.28% to 77.98%. Then, the highest accuracy model is used to obtain the vector representation of extended candidate words and seed words, and we use those vectors to calculate cosine coefficient between seed words and candidate words. Some results are shown in the figure below.

Table 3. Some result of cosine coefficient

| seed words | candidate words | cosine |
|---|---|---|
| Central economic work conference | Foreign trade | 0.998631 |
| | Firmly pursuit | 0.998383 |
| | Year | 0.998361 |
| | Next year | 0.998355 |
| | Enhance | 0.998339 |
| Local debt | Fund | 0.997658 |
| | Loss | 0.997426 |
| | Project | 0.996735 |
| | Unified | 0.996544 |
| | Lack of funds | 0.996414 |

|  | City | 0.994121 |
|  | Rail | 0.984592 |
| Railway | Base | 0.977009 |
|  | Sale | 0.976308 |
|  | Operate | 0.975765 |

In order to expand the indicator words under each secondary indicator, we put together all the expanded candidate words generated by the indicator words under each secondary index, and selects the candidate words with the top 50 similarity for further screening, removing adjectives, words without Baidu Index and words with ambiguous meaning, and finally obtains the expanded entries of each secondary indicator. After the expansion, the index system has a total of 93 indicator words. The final expansion results are shown in Table 4:

Table 4. Indicator system after expansion

| Primary indicators | Secondary indicators | Indicator entries |
|---|---|---|
| Government investment | Government economic attitude | Central economic work conference, Local debt, **Foreign trade**, **Fiscal policy**, **Macroeconomic, Globalization, Carbon peak** |
|  | Infrastructure investment | Infrastructure, Affordable housing, Highway, Railway, Airport, Electrified wire netting, Freeway, Energy saving and emission reduction, Livelihood project, Water conservancy, **Freight transport**, **Communal facilitie**, **Travel**, **Farming and forestry**, **Dilapidated house transforming, Social security** |
|  | New infrastructure investment | New infrastructure, 5G, UHV, New energy vehicles, Industrial Internet, Artificial intelligence, Data center, **Man-machine**, **Engineering construction, Internet, Intelligence, Medical treatment, Innovation, Machine** |
| Investment influenced by government | State-owned enterprise investment | State-owned enterprise, Central enterprises, Public institutions, **Finance, organization, Structural reform, Central government, Administration** |
| Private investment | Real estate investment | Real estate, Housing price, Real estate policy, **Housing, residence, lease, House, market economy, Building** |

| | Non real estate investment | Limited company, Private enterprise, Self-employed laborer, Enterprise Registration, Office Building, Shops, Factory, Workshop, **Machining, Centre, Technician, Modernization** |
|---|---|---|
| Foreign investment | Foreign investment | Fdi, Foreign enterprise, Joint venture, The open policy, **Gdp, Long-term investment, Mainland, Imf, US Department of Commerce** |
| Investment environment | Industry aspect | Real economy, Manufacturing, Industry, Small and micro businesses, Industrial upgrading, **Bank, Transformation, Handling** |
| | Economic aspect | Tax reduction, Interest rate cut, Reserve requirement ratio cut, Loan, Interest rate, **Consumption, Enterprise, People's Bank of China, Discount, Auction** |

## *5.3. Analysis investment activity index in China*

Web search data can make up for the problems of traditional data that are not timely and have limited sources, but Web data also contains a lot of invalid information, which may affect the accuracy of index measurement. Therefore, based on China's fixed asset investment, this paper calculates the Pearson correlation coefficient between the above 93 indicators and the investment, removes the indicators whose absolute value of the correlation coefficient is less than 0.1, and the indicators whose Baidu Index is shorter than the research time of this paper, and obtains 66 indicators after screening. In the process of index expansion, more words related to investment activity have been included in the indicator system, and more information related to investment has also been included in theory. Therefore, the expanded indicator words should have better ability to explain the amount of investment in fixed assets. In this paper, the indicators after correlation screening and the expanded indicators after the same correlation screening are linearly regressed with China's fixed asset investment. In order to avoid the problems that may be caused by too many variables, this paper uses factor analysis to divide five factors based on primary indicators, and calculates the factor scores of five primary indicators before and after the expansion as independent variables. The regression results are shown in Table 5:

Table 5. Regression results before and after expansion

|  | R square | Adjust R square | F-value | P value of F-test |
|---|---|---|---|---|
| **Before expansion** | 0.148 | 0.059 | 1.662 | 0.162 |
| **After expansion** | 0.204 | 0.121 | 2.463 | 0.046 |

It can be found that the R-square and p value of F-test of the indicator factors are greatly improved after expansion, indicating that through the indicator expansion based on semantic similarity, we have successfully obtained more indicators containing information about fixed asset investment, which is positive for the calculation of investment activity index.

Next, we use the entropy weight method to calculate the weight of each indicator. When we use entropy weight, the indicator is divided into positive and negative. Considering the relationship between the indicator entries and China's investment activity, this paper divides "US Department of Commerce" into negative indicators, because the China-U.S. trade war has had a negative impact on China's economy. In addition, considering the special status of real estate investment in China and the government's attitude towards the stability of the real estate industry, this paper sets "real estate", "house price", "residence", "house" and "building" as two-way indicators, that is, when index larger than median, it is treated as a negative indicator. And when index less than median, it is treated as a positive indicator.

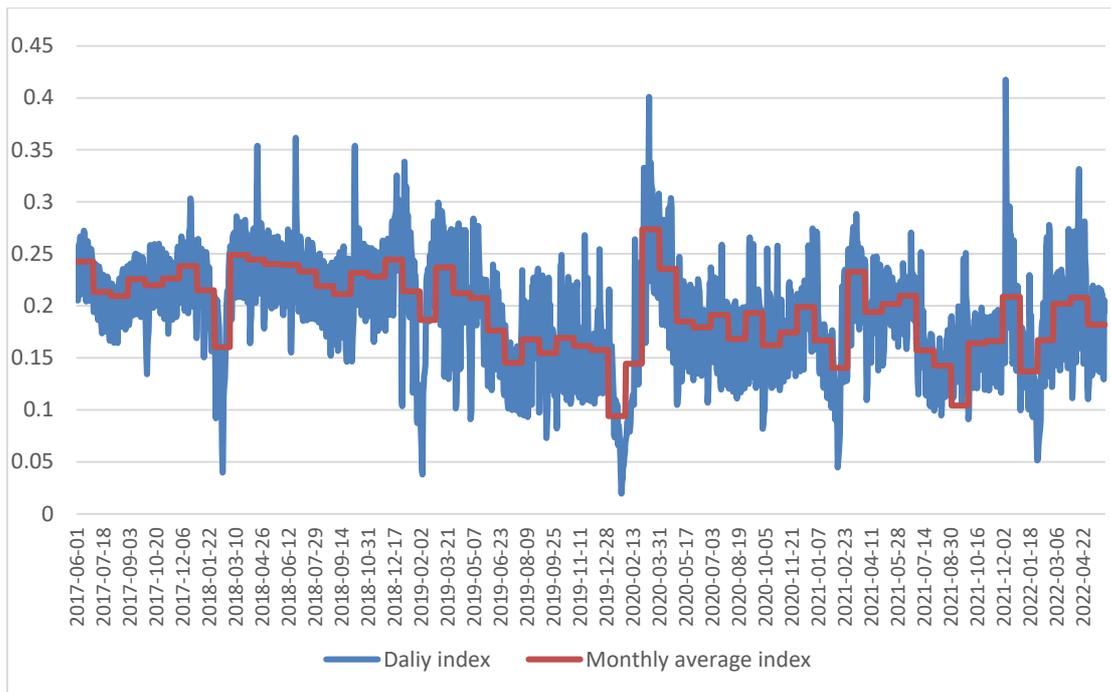

Figure 4. Investment activity index in China

Figure 4 shows the daily index and monthly average index of China's investment activity. We can find that: first, the investment activity is greatly affected by the lunar new year. The Chinese Spring Festival is generally in January or February of the Gregorian calendar. The investment activity decreases significantly in January and February of each year. After March, production starts to recover, and the investment activity is also at high level in a year. Second, in 2019, the investment activity showed a downward trend. In 2019, China's GDP growth rate was only 6%, down 0.7 percentage points from 6.7% in 2018, while the GDP growth gap in 2016, 2017 and 2018 was within 0.2%. In addition to the decline in economic growth, the United States has imposed tariffs on Chinese goods since the beginning of 2019, which has hit China's exports. Under the influence of various negative factors, China's investment activity continued to decline in 2019. Third, the outbreak of COVID-19 has greatly affected the activity of investment in China. COVID-19 broke out in Wuhan in early 2020 and then spread in many provinces of China. At the beginning of 2020, China's investment activity also fell to the lowest level. After the epidemic was basically controlled, the investment activity recovered explosively. Fourth, from the second half of 2021, the investment activity has decreased and the fluctuation has increased. In the first half of 2021, the investment activity is higher than that in 2019, but in the second half of 2021, the activity has decreased and fluctuated greatly. The investment activity at the beginning of 2022 is lower than that in the past, and the prospect is not optimistic.

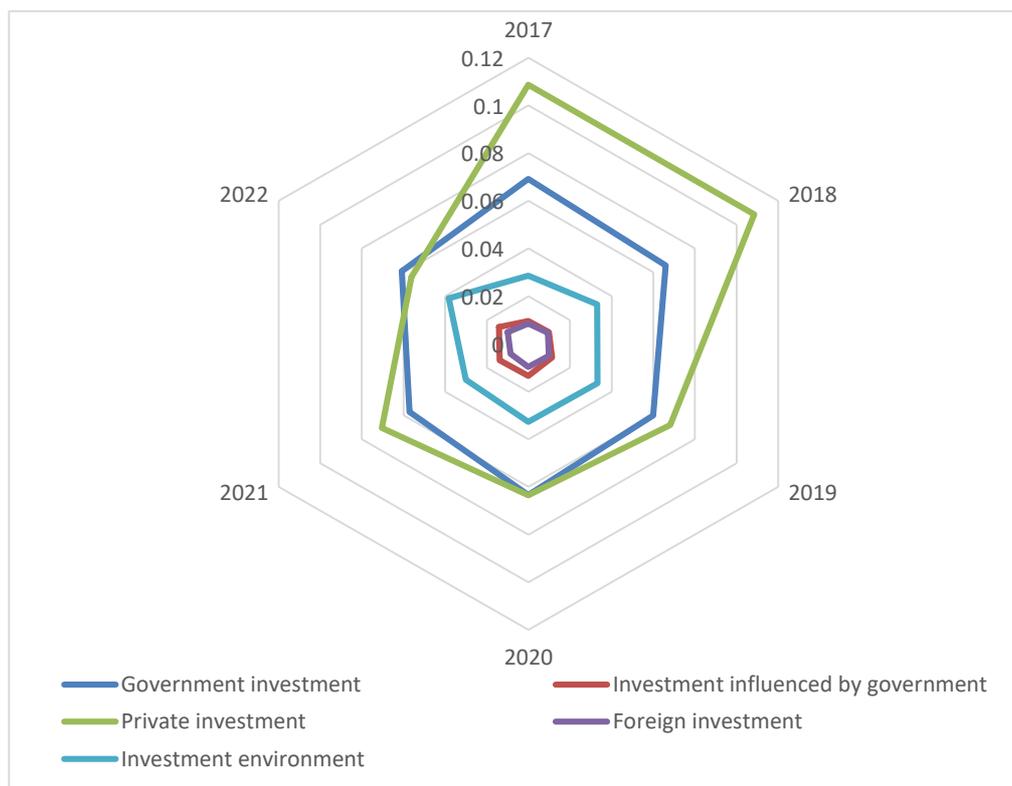

Figure 5. Annual average Investment activity index in 5 dimension from 2017-2022

Figure 5 shows the annual average index value of the five primary indicator dimensions of China's investment activity from 2017 to 2022. From the perspective of private investment, the activity of private investment has been at the highest until 2022. It was at a high level in 2017 and 2018. It decreased a lot in 2019 and continued to decline in 2020. It rebounded in 2021 and fell to the lowest level in 2022. This reflects that the willingness of Chinese private investment continues to decline after the China- U.S. trade war and the slowdown of China's economic growth, and is further reduced by the impact of the COVID-19. From the time of the outbreak of the COVID-19, novel coronavirus was first discovered in China in early 2020. In December 2021, the Omicron virus entered the Chinese mainland for the first time. Compared with the activity in 2020 and 2022, it can be seen that although the symptoms of Omicron virus infection are mild and some foreign countries have resumed production, the impact on private investment will be greater after the wide spread of Omicron virus in China than the outbreak in 2020. In the dimension of government investment, most of the time the activity of government investment is second, but in 2022 government investment became the most active dimension. From 2017 to 2019, the activity of government investment decreased year by year, and increased in 2020 when the epidemic broke out. Compared with the change of private investment activity, it can be found that after 2020, the activity of government investment and private investment changed in reverse. The change of investment dimension influenced by the government is different from other dimensions. It has maintained growth from 2017 to 2022. On the one hand, it shows that state-owned enterprises and central enterprises have a good development momentum in recent years and maintained growth; On the other hand, it also shows that the investment behavior of the Chinese government is more completed by state-owned enterprises, central enterprises and other state-controlled enterprises. The dimension of investment environment has a prominent growth in 2022, which indicates that China's investment environment in 2022 shows obvious signs of easing.

Table 6. Annual average Investment activity index of provinces in Chinese mainland from 2017-2022

| year | 2017 | 2018 | 2019 | 2020 | 2021 | 2022 |
|---|---|---|---|---|---|---|
| Shandong | 0.2890 | 0.2944 | 0.2608 | 0.2595 | 0.2579 | 0.2655 |
| Guizhou | 0.2892 | 0.2926 | 0.2904 | 0.2915 | 0.2771 | 0.2889 |
| Jiangxi | 0.2836 | 0.3094 | 0.2774 | 0.2924 | 0.2868 | 0.3062 |
| Chongqing | 0.3072 | 0.3236 | 0.2983 | 0.2898 | 0.2838 | 0.2957 |
| Neimenggu | 0.2813 | 0.2959 | 0.2860 | 0.3046 | 0.2948 | 0.3075 |

| Province | | | | | | |
|---|---|---|---|---|---|---|
| Hubei | 0.3090 | 0.3220 | 0.2957 | 0.2767 | 0.2781 | 0.2850 |
| Liaoning | 0.2928 | 0.3156 | 0.2846 | 0.2998 | 0.2864 | 0.2949 |
| Hunan | 0.3148 | 0.3273 | 0.2994 | 0.2961 | 0.2980 | 0.3146 |
| Fujian | 0.3032 | 0.3141 | 0.2803 | 0.2771 | 0.2763 | 0.2862 |
| Shanghai | 0.3028 | 0.3091 | 0.2778 | 0.2599 | 0.2585 | 0.2553 |
| Beijing | 0.3036 | 0.3088 | 0.2770 | 0.2627 | 0.2554 | 0.2489 |
| Guangxi | 0.2981 | 0.3162 | 0.3002 | 0.3155 | 0.3042 | 0.3160 |
| Guangdong | 0.2721 | 0.2770 | 0.2544 | 0.2428 | 0.2291 | 0.2289 |
| Sichan | 0.3022 | 0.3114 | 0.2775 | 0.2838 | 0.2649 | 0.2672 |
| Yunnan | 0.2937 | 0.3197 | 0.3025 | 0.2951 | 0.2876 | 0.2958 |
| Jiangsu | 0.2832 | 0.2953 | 0.2653 | 0.2554 | 0.2592 | 0.2611 |
| Zhejiang | 0.2844 | 0.2923 | 0.2682 | 0.2632 | 0.2609 | 0.2580 |
| Qinghai | 0.1835 | 0.2042 | 0.2007 | 0.2063 | 0.2066 | 0.2263 |
| Ningxia | 0.1899 | 0.2056 | 0.2151 | 0.2288 | 0.2288 | 0.2578 |
| Hebei | 0.2958 | 0.3085 | 0.2728 | 0.2974 | 0.2813 | 0.2879 |
| Heilongjiang | 0.2578 | 0.2791 | 0.2718 | 0.2966 | 0.2747 | 0.2861 |
| Jilin | 0.2951 | 0.3030 | 0.2857 | 0.2908 | 0.2890 | 0.2936 |
| Tianjin | 0.3027 | 0.3243 | 0.3012 | 0.2960 | 0.2945 | 0.2966 |
| Shaanxi | 0.3102 | 0.3232 | 0.3047 | 0.2967 | 0.2932 | 0.3028 |
| Gansu | 0.2615 | 0.2737 | 0.2676 | 0.2853 | 0.2698 | 0.3090 |
| Xinjiang | 0.2547 | 0.2838 | 0.2613 | 0.2705 | 0.2835 | 0.3012 |
| Henan | 0.2788 | 0.2976 | 0.2608 | 0.2573 | 0.2565 | 0.2618 |
| Anhui | 0.2832 | 0.3038 | 0.2749 | 0.2899 | 0.2857 | 0.2956 |
| Shanxi | 0.2919 | 0.3082 | 0.2991 | 0.3118 | 0.2951 | 0.3129 |
| Hainan | 0.2393 | 0.2626 | 0.2557 | 0.2752 | 0.2741 | 0.2978 |
| Xizang | 0.1478 | 0.1595 | 0.1614 | 0.1647 | 0.1725 | 0.1748 |

We further concern about the changes in the investment activity of provinces in Chinese Mainland. From 2017 to 2022, the investment activity of provinces changed greatly. From 2017 to 2018, more provinces became active in investment, but the number of provinces with high activity decreased significantly in 2019, only central provinces remained active. In 2020, some northern and southern provinces became more active. In 2021, the number of provinces with high activity continued to fall. In 2022, the number of provinces with high activity was higher. However, since we only use the data of the first five months of 2022 to calculate the average value, and March and April are usually the periods with the highest activity in a year, the activity index in 2022 is overestimated compared with the average value of activity index in other years. Therefore, for the investment activity index in 2022, we only focus on the value between provinces.

More specifically, investment activity of economically developed provinces such as Jiangsu, Zhejiang and Guangdong declined after 2019, which explains the weakness of private investment from the perspective of provinces, while Guangxi and Hunan have remained active in investment. It is worth noting that before 2019, Beijing's investment activity had been higher than surrounding provinces, but after 2020, Beijing's investment activity declined seriously and was lower than surrounding regions. Shanghai also shows a similar situation. Generally speaking, the provinces with high investment activity in China after 2019 are mainly economically underdeveloped provinces, which are concentrated in the north and South; The investment activity of developed provinces and Megacities, such as Beijing, Shanghai, Guangdong, Jiangsu and Zhejiang, has declined, which may be due to the relatively low effect of investment on the economy due to their developed economies. With the economic downturn, number of jobs in developed provinces may decrease, leading to more people returning to their home provinces. We expect that under the condition of high government investment activity, the difference in investment activity among provinces will continue to exist, or even further expand.

After analyzing the changes of China's investment activity index, we continue to consider the relationship between investment activity and investment in fixed assets. Theoretically, the investment idea takes time to be done, so the investment in fixed assets lags behind the activity index. The length of lag period may vary in different economic environments. Therefore, we have calculated the correlation between China's investment activity index and fixed asset investment before 2020 and after June 2020, taking the outbreak of COVID-19 in early 2020 as the boundary. Since the data of China's fixed asset investment in January of each year are not published, but only the cumulative amount from January to February, we have adjusted the investment activity index accordingly.

Table 7. Correlation coefficient between investment activity index and investment in fixed assets before and after COVID-19

| Lag | 0 month | 1 month | 2 months | 3months | 4 months | 5 months |
|---|---|---|---|---|---|---|
| Before | -0.1147 | 0.1303 | 0.1197 | 0.3657 | 0.7754 | 0.1713 |
| After | 0.1319 | 0.2888 | 0.1225 | 0.1839 | 0.2499 | -0.1217 |

Through the correlation coefficients of different lag periods, we can find that the lag period with the highest correlation before COVID-19 is 4 months, while lag period with the highest correlation after COVID-19 is 2 months. This may indicate that under the outbreak of the COVID-19 and the expectation of economic downturn, investors decided to reduce investment with a long construction period, thereby reducing uncertainty. As cities will be put on lockdown after finding new coronavirus infected persons, the investment with a long construction period is faced with the risk of interruption of the construction process and failure of getting return on time, which may lead to serious consequences of broken capital chain for small and medium-sized enterprises. However, due to the short transmission time of Omicron virus in Chinese Mainland, we may need more long-term data to draw a conclusion.

**6. Conclusion**

This paper uses natural language processing technology to creatively construct an index that can reflect China's investment activity in a timely manner, and systematically analyzes the variation characteristics of China's investment activity in various dimensions and different provinces, and draws the following conclusions and corresponding enlightenment:

First, Investment activity in China declined in 2019, rebounded in 2020 after the impact of COVID-19, but declined again in late 2021 and 2022. This indicates that China's investment activity began to decline after 2019. There was a rebound period after the outbreak of COVID-19, but the decline continued, and government investment gradually played a major role in this process. Second, the private investment activity continues to decline except 2021, while the government investment activity increases, and has surpassed the private investment activity in 2022. In addition, the investment environment appears to be relatively relaxed in 2022. The decline in private investment is hard to stop, so The Chinese government should continue to increase government-led investment. Third, after 2019, the number of provinces with high investment activity in China has gradually decreased. The investment activity of economically developed provinces has declined more, while the investment activity of some relatively underdeveloped provinces in the north and south of China has been relatively high. As for all provinces, investment in underdeveloped provinces plays an obvious role in driving the economy, so we should continue to strengthen investment in these provinces to maintain economic stability. Fourth, the lag between fixed asset investment and investment activity has shortened before and after COVID-19. It means that frequent lockdowns and quarantine policies have had a negative impact on investors' long-term investment, and the Chinese government should improve epidemic prevention measures to reduce the impact of the epidemic on investment activity.

**Disclosure statement**

No potential conflict of interest was reported by the authors

**Funding**

This work of Xiaobin Tang was supported by National Social Science Foundation of China (No. 22&ZD164), National Social Science Foundation of China(Grant Number No. 17BJY055), Outstanding Yong Scholars Funding Program Of UIBE (No. 21JQ09), CAAI-Huawei MindSpore Open Fund (No.CAAIXSJLJJ-2021-045A)